\documentclass[letterpaper, 10pt, conference]{ieeeconf}
\IEEEoverridecommandlockouts
\usepackage{amsmath,amssymb,amsfonts}
\usepackage{algorithmic}
\usepackage{graphicx}
\usepackage{xcolor}
\usepackage{cite}
\usepackage{graphicx}      
\usepackage{amsmath}
\usepackage{amssymb}
\usepackage{multirow}
\usepackage{makecell}
\usepackage{booktabs}
\usepackage{color}
\usepackage[ruled]{algorithm2e}

\usepackage{url}

\newcommand{\revision}{\color{black}}
\newcommand{\revise}{\color{black}}

\IEEEoverridecommandlockouts                              
\title{\LARGE\bf Unlocking Energy Flexibility From Thermal Inertia of Buildings:\\ A Robust Optimization Approach} 

\author{Yun Li, Neil Yorke-Smith, and Tamas Keviczky
\thanks{The work was supported by the Brains4Buildings project under the Dutch grant programme for Mission-Driven Research, Development and Innovation (MOOI).}
\thanks{Y. Li and T. Keviczky are with Delft Center for Systems and Control, Delft University of Technology, The Netherlands
{\tt\small y.li-39@tudelft.nl; T.Keviczky@tudelft.nl}}
\thanks{N. Yorke-Smith is with the STAR Lab, Delft University of Technology, The Netherlands
{\tt\small n.yorke-smith@tudelft.nl}}
}

\begin{document}
\maketitle
\thispagestyle{empty}
\pagestyle{empty}

\begin{abstract}
\revision Towards integrating renewable electricity generation sources into the grid, an important facilitator is the energy flexibility provided by buildings' thermal inertia.
Most of the existing research follows a single-step price- or incentive-based scheme for unlocking the flexibility potential of buildings. In contrast, this paper proposes a novel two-step design approach
for better harnessing buildings' energy flexibility.
In a first step, a robust optimization model is formulated for assessing the energy flexibility of buildings in the presence of uncertain predictions of external conditions, such as ambient temperature, solar irradiation, etc. In a second step, energy flexibility is activated in response to a feasible demand response (DR) request from grid operators without violating indoor temperature constraints, even in the presence of uncertain external conditions. The proposed approach is tested on a high-fidelity {\tt Modelica} simulator to evaluate its effectiveness. Simulation results show that, compared with price-based demand-side management, the proposed approach achieves greater energy reduction during peak hours.
\end{abstract}

\section{Introduction}
\label{sec:intro}

Since the turn of the 21st century, more and more renewable energy sources (RES) are integrated into energy networks. While the increased penetration of RES is beneficial for the decarbonization of energy systems, it also brings challenges to energy systems in terms of greater volatility, intermittency and uncertainty, and lesser controllability. These unfavorable factors make it difficult to guarantee the balance between energy demand and supply at all times, which is important for energy networks.

To hedge against the influences caused by RES and to stabilize energy systems, the concept of demand-side management has been extensively investigated in the recent literature. Unlike supply side management, which involves adjusting the energy generation and supply according to the need of energy consumers, demand-side management refers to the modification of end users' energy demand and the exploitation of demand flexibility of energy consumers, which is usually achieved via demand response (DR) strategies. According to the Clean Energy Package of the European Union, utilizing end-user flexibility plays an important role in supporting the decarbonization of the energy system \cite{Abl:45}. 

As a major energy consumer, buildings contribute to roughly 40\% percent of the total energy consumption; the energy consumed in buildings for heating, ventilation and air-conditioning (HVAC) accounts for about half of the total energy consumption of buildings in Europe \cite{perez2008review,bianchini2016demand}.  In addition, according to the European Energy Performance of Buildings Directive (EPBD), most of the utility buildings will be equipped with a building automation and control system from 2026
onwards \cite{EPBD3}. As a result, the increasing availability of a large number of controllable heating devices, smart metering and advanced computation technology makes it possible to improve building energy management strategies to unlock the energy flexibility of buildings. This flexible energy can then promote the decarbonization of energy systems, e.g., solving grid congestion, which currently is a challenging issue in the Netherlands.


Among the three main types of flexibility sources - buildings' thermal inertia, heat carriers, and heat storage devices - the flexibility emanated from buildings' thermal inertia is the most accessible and is fast in response to provide intra-day energy balancing, but with limited capacity \cite{golmohamadi2021optimization,golmohamadi2022integration,vandermeulen2018controlling}. This type of flexibility is reflected in the buildings' thermal dynamics and can be unlocked by actively over-heating/under-heating buildings \cite{kensby2015potential}. In order to effectively harness this  flexibility, we argue that advanced control algorithms that can consider external signals such as climate information, electricity prices and so on, and utilize the thermal mass of buildings should be designed to replace conventional rule-based controllers.

Model predictive control (MPC) is regarded as a promising advanced control strategy for building control systems due to its versatility in considering external influences and formulating dynamic economic cost and system constraints as well as other factors. MPC-based building control systems are extensively studied in the recent literature. In general, most of the MPC-based DR strategies can be classified into two categories: 1) \textit{price-based program} \cite{oldewurtel2013towards,golmohamadi2021optimization,d2019mapping,wu2018bi}, which means the building management system (BMS) voluntarily adjusts its energy consumption in reaction to certain economic signals, e.g., electricity prices, and 2) \textit{incentive-based program} \cite{bianchini2016demand}, which means the BMS reduces its energy consumption according to DR requests that specify a profile of energy consumption during a certain time period with corresponding incentives. More details of the two categories of DR programmes can be found in  \cite{golmohamadi2022integration} and \cite{d2019mapping}. 

{\revision
It should be pointed out that a common limitation for both price- and incentive-based programmes is that the expected profile of energy consumption might not be achieved, either because the BMS is not price-responsive or the DR request is too aggressive for the BMS. Even if the expected profile of energy consumption for buildings is achieved, it is possible that the flexibility potential of buildings is not fully exploited. The main reason for these limitations is that the above-mentioned approaches fail to incorporate an explicit description of the flexibility potential of buildings. While some works have investigated the problem of energy flexibility quantification and exploitation for buildings, e,g., \cite{pan2017feasible,vrettos2016robust,qureshi2018hierarchical,gorecki2017experimental}, there is a lack of unified framework for unlocking buildings' energy flexibility and concrete procedures for formulating and solving the problem of flexibility assessment and exploitation.

In addition, buildings are exposed to the influence of many exogenous disturbances, such as ambient temperature, solar irradiation, occupancy patterns. If such factors are not properly considered in the control design, it may result in occupant discomfort, low operational efficiency, and high energy cost \cite{zhang2013scenario,ostadijafari2020tube,shi2021robust}. However, existing works on energy flexibility assessment of buildings, such as \cite{pan2017feasible,vrettos2016robust,qureshi2018hierarchical,gorecki2017experimental}, merely consider deterministic control design, and the issue of uncertainties has not gained sufficient attention.

Motivated by the above research gap, this paper aims at providing a unified framework for investigating the energy flexibility associated with buildings' thermal inertia in the presence of uncertain external conditions. 
Our main contributions and innovations are summarized as follows:
\begin{itemize}
\item Unlike existing price- or incentive-based approaches, a novel two-step design framework is proposed to more efficiently exploit the energy flexibility of buildings for demand-side management.
\item A robust optimization (RO) problem is formulated for assessing the energy flexibility of buildings in the presence of uncertain predictions of external conditions. Duality of linear programming is applied to make the RO problem computationally tractable. Solutions for reducing online computational burden are also discussed.
\item Numerical experiments are implemented on a {\tt Modelica} building simulator to test the effectiveness of the proposed scheme. Compared with the price-based program for demand-side management, the proposed approach can better unlock the energy flexibility of buildings and achieve more energy reduction during peak hours.
\end{itemize}
}
The following structure is adopted. Section~\ref{sec:problem_formulation} describes the approaches for obtaining a control-oriented model, and introduces the proposed design framework. Section~\ref{sec:flexibility_assessment} elaborates the mathematical formulation of the proposed scheme with in-depth discussions. Section~\ref{sec:simulation} gives extensive numerical simulation results. Section~\ref{sec:conclusion}  concludes this paper and points out some options for further development.

\section{Building Thermal Dynamics and Problem Formulation}\label{sec:problem_formulation}
In order to investigate the energy flexibility provided by building thermal dynamics, a mathematical model describing building thermal dynamics is required. Among various modeling techniques, RC-network based approaches are of particular interest to building control community since the derived control-oriented model usually gives satisfactory prediction performance while retaining tractable computational complexity. More details about RC-network based modelling for building thermal dynamics can be found in \cite{bacher2011identifying}. Besides RC-network models, our proposed approach can also be easily extended to utilize the autoregressive models of building thermal dynamics in \cite{madsen2015thermal}.

Without loss of generality, we assume that the building thermal dynamics is developed via RC-network approach, and consider electricity as heating/cooling power sources with constant coefficient of performance for heating/cooling devices. Then, the evolution of room temperature can be given as \cite{shi2021robust,bianchini2016demand,vrettos2016robust}:
\begin{equation}\label{eq:dynamics}
    x_{k+1} = Ax_k + B u_k + Rw_k + Dd_k
\end{equation}
where $x_k\in\mathbb{R}^n$ is the state vector including indoor temperature and building envelop temperature; $u_k\in\mathbb{R}^m$ is the heating/cooling power provided by the local RES; $w_k\in\mathbb{R}$ is the power drawn from the grid; $d_k\in\mathbb{R}^p$ is the uncontrolled external inputs, such as ambient temperature and solar radiation; the subscript $k$ denotes the $k$-th time instant, and $k=0$ denotes the initial time step. 

It should be mentioned that while electricity is assumed as the heating/cooling power in our design, the proposed approach is independent of the heating/cooling technologies and resources, and can be extended to buildings with different types of heating/cooling systems easily.

As discussed in Section~\ref{sec:intro}, most of the existing approaches in the literature investigating the energy flexibility of buildings follow the so-called price-based or incentive-based program, which is a one-step approach. Specifically, the BMS either adjusts its energy consumption according to some price signals or follows a specific energy consumption profile designated via DR requests. {\revise However, these DR programmes can be conservative since they do not take into account the operation situation of buildings when more flexibility is needed, as shown in our simulation results Fig. \ref{fig:4}. On the one hand, it is possible that the BMS might not be sufficiently price-responsive to fully exploit the flexibility potential of buildings.} On the other hand, the DR request sent by grid operators might be too aggressive to be achieved by the BMS. As a result, the flexibility potential of buildings is not fully activated, and the expected DR requests are not achieved.

\begin{figure}[tb]
    \centering
    \includegraphics[width = 0.9\linewidth]{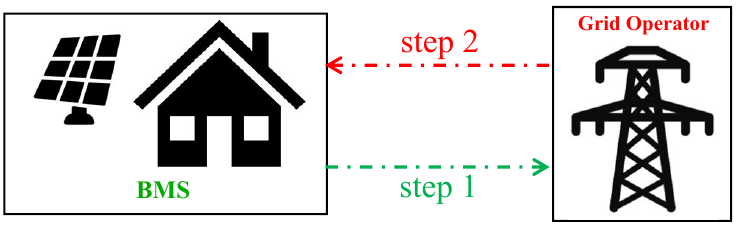}
    \caption{Diagram of the proposed scheme: step 1) flexibility assessment by BMS, step 2) flexibility exploitation through DR by grid operator. }
    \label{fig:communication_structure}
\end{figure}

{\revision
We hereby propose a new unified design framework for assessing and exploiting the energy flexibility of buildings. In the proposed design framework, the building model itself, local RES, and the energy network are considered. Our demand-side management comprises two steps: (1) in the first step, a robust formulation is proposed for quantitatively assessing the flexibility potential emanating from buildings' thermal inertia while considering uncertain external conditions, and the information of flexibility capacity is sent to grid operators; (2) in the second step, flexibility is exploited via the format of judicious DR requests, which specify profiles of energy consumption during a given time slot, that are compatible with the amount of available energy flexibility. 

A schematic diagram of the proposed scheme is depicted in Fig. \ref{fig:communication_structure}. With the proposed design framework, the flexibility potential of buildings can be robustly assessed and exploited without sacrificing indoor temperature comfort even in the presence of uncertain external conditions. In addition, since the task of flexibility assessment for buildings is performed by BMS, the workload of the grid operator for demand-side management is reduced, and the privacy information of buildings is secured. {\revise In addition, as explained in \textit{Remark 3}, for achieving this bidirectional interaction between power grid and BMS, our proposed approach actually does not cause remarkably increased communication burden.}
}



\section{Flexibility Assessment and Control Strategy}\label{sec:flexibility_assessment}
The main challenge for the proposed approach is quantitatively assessing the flexibility potential. In this section, the mathematical formulations for assessing the energy flexibility of buildings are elaborated. First, we introduce the metrics for quantitatively describing the flexibility. Then, based on these metrics, we present the detailed formulations for assessing the energy flexibility.

\textit{Notation}: uppercase letters denote matrixes, bold-face lowercase letters are stacked time sequences of relevant signals. $\mathbb{R}^d$ and $\mathbb{R}_+$ denote $d$-dimensional real space and non-negative real scalar, respectively. $\mathbb{B}^{x\times y}$ denotes binary matrix with $x$ by $y$ dimension. $[\cdot]_i$ denotes the $i$-th row of a given matrix or the $i$-th element of a given vector.

Based on the system dynamics in \eqref{eq:dynamics}, we have the following time-lifted representation of the room temperature prediction over $N$ time steps:
\begin{equation}\label{eq:lifted}
    {\bf x} = F_xx_0 + F_u{\bf u} + F_w{\bf w} + F_d{\bf d}
\end{equation}
where ${\bf x} = [x_1^{\mathrm{T}},\cdots,x_N^{\mathrm{T}}]^{\mathrm{T}}\in\mathbb{R}^{nN}$, ${\bf u} = [u_0^{\mathrm{T}},u_1^{\mathrm{T}},\cdots,u_{N-1}^{\mathrm{T}}]^T\in \mathbb{R}^{m N}$, ${\bf w} = [w_0,\cdots,w_{N-1}]^{\mathrm{T}}\in\mathbb{R}^{N}$, ${\bf d} = [d_0^{\mathrm{T}},\cdots, d_{N-1}^{\mathrm{T}}]^{\mathrm{T}}\in\mathbb{R}^{pN}$. The definitions of $F_x$, $F_u$, $F_w$ and $F_d$ are available in MPC design references, e.g., \cite{parsi2022computationally}.

While there is no such commonly accepted definition of the flexibility potential of buildings, it can be summarized from the existing literature, especially the report of the IEA EBC Annex 67 \cite{annex67}, that there are three main parameters for describing the energy flexibility: capacity, duration, and starting time instant.

The power from the grid is decomposed into two parts:
\begin{equation}\label{eq:grid_power}
    {\bf w} = {\bf \bar{w} } + M\tilde{\bf w}
\end{equation}
where ${\bf{\bar w}} = [\bar{w}_0,\cdots,\bar{w}_{N-1}]^{\mathrm{T}}\in\mathbb{R}^N$ denotes the nominal power drawn from the grid, which can be determined by the BMS according to day-ahead electricity prices; $\tilde{\bf w} = [\tilde{w}_{0},\cdots,\tilde{w}_{h-1}]^{\mathrm{T}}\in \mathbb{R}^{h}$ is the flexible power consumption that will be determined by the grid operator in the format of DR requests, and $h$ is the length of the duration of flexibility. Note that since the exact sequences of flexible power consumption $\tilde{\bf w}$ will be designated by the grid operator, it is a source of uncertainty for the BMS.

\textit{Remark 1}: The matrix $M\in\mathbb{B}^{N\times h}$ is for specifying the time period during which the flexibility is assessed. In our design, we assume that the BMS and the grid operator come to an agreement for determining the activation period of flexibility, and hence $M$ and $h$ are determined accordingly before assessing the energy flexibility. Note that the elements and dimension, namely $h$, of $M$ can also be decision variables to be optimized in the process of flexibility assessment, and the resulting flexibility assessment formulation becomes a mixed-integer program. For example, the optimization problem can be formulated to maximize the duration of the flexible power consumption period.

In this work, we focus on using the energy flexibility from buildings to solve the grid congestion and peak shaving problem, which is currently a challenging issue in the Netherlands, and hence we consider the case that the BMS reduces its power consumption from the grid. Without loss of generality, the proposed scheme can be easily extended to the case of positive flexible energy adjustment. The capacity of the flexibility $\tilde {\bf w}$ is described by the following two aspects: amplitude and ramping rate
\begin{subequations}
    \begin{align}
       & -\gamma_{1} \leq \tilde w_t \leq 0, \ t = 0,1,\cdots,h-1\label{eq:magnitude}\\
       & -\gamma_{2} \leq \tilde w_{t+1} - \tilde w_{t} \leq \gamma_{2},\ t = 0,\cdots,h-2\label{eq:ramp_rate}
    \end{align}
\end{subequations}
where $\gamma_{1}\in\mathbb{R}_+$ and $\gamma_{2}\in\mathbb{R}_+$.
Then, the feasible region of $\tilde{\bf w}$ can be compactly expressed with the following linear constraint
\begin{equation} \label{eq:flexi_set}
\mathcal{W} := \left \{ \tilde{\bf w} \mid H_w\tilde {\bf w} - g_w \leq 0 \right\}
\end{equation}where $H_w\in\mathbb{R}^{(4h-2)\times h}$, and $g_w \in\mathbb{R}^{4h-2}$. Particularly, note that $g_w$ contains $\gamma_{1}$ and $\gamma_{2}$, which we intend to use as decision variables to be determined in flexibility assessment.

The above formulation specifies the amplitude and ramping-rate constraints of flexible power consumption $\tilde{\bf w}$, which together with the matrix $M$ are sufficient to quantitatively describe the energy flexibility in terms of the basic parameters. 

To make the designed approach more practical, the uncertainties in the prediction of uncontrolled exogenous input $\bf d$, including solar radiation, ambient temperature, internal gains and etc, are also considered in our design. As a result, another source of uncertainty for BMS is the prediction error of external input, which is defined as $\tilde{\bf d} := {\bf d} - {\bf \hat d} = [\tilde{d}^{\mathrm{T}}_0,\cdots,\tilde{d}^{\mathrm{T}}_{N-1}]^{\mathrm{T}}$. Since in practice most sets of thermal uncertainties, such as ambient temperature, solar irradiation and internal heat gain, can be denoted as hyper-rectangles, without loss of generality $\tilde {\bf d}$ is assumed to be restricted to the following polytopic set 
\begin{equation}\label{eq:predict_uncertain}
\mathcal{\tilde{D}} := \{{\bf \tilde{d}}\mid H_d{\bf \tilde{d}} - g_d\leq 0\}
\end{equation}with $H_d\in\mathbb{R}^{l_d\times pN}$ and $g_d\in\mathbb{R}^{l_d}$.

The design objective for energy flexibility assessment is to explore the maximal scope of the flexible power consumption $\tilde{\bf w}$ whilst guaranteeing room temperature within its comfort band. 
{\revision In our design, we assume that only the power from local RES can be controlled in real-time by BMS. The feasible set of energy consumption from local RES is 
\begin{equation}\label{eq:feasible_input}
\mathcal{F}(x_0,\mathbf{w},\mathbf{d}) :=\! \left\{
\mathbf{u}\left|
\begin{aligned}
&\ {\bf x} = F_xx_0 + F_u{\bf u} + F_w{\bf w} + F_d{\bf d}\\
&\ \mathbf{w} = \bar{\mathbf{w}} + M\tilde{\mathbf{w}},\ \mathbf{d} = \mathbf{\hat{d}} + \tilde{\mathbf{d}}\\
&\ \mathbf{x}\in\mathcal{X},\mathbf{u}\in\mathcal{U}\\
&\ \forall \tilde{\mathbf{w}}\in\mathcal{W},\ \forall \tilde{\mathbf{d}}\in\tilde{\mathcal{D}}
\end{aligned}
\right.
\right\}
\end{equation}
where $\mathcal{X}$ and $\mathcal{U}$ are feasible sets of room temperature and local RES, respectively. {\revise Since considering arbitrary control policy for $\bf u$ will make the resulting optimization problem computationally intractable, to balance the computational burden and the optimality of the proposed approach}, we adopt the following affine control policy, which is an affine decision rule w.r.t. the uncertainties $\tilde{\bf w}$ and $\tilde{\bf d}$
\begin{equation}\label{eq:LDR}
    {\bf u} = {v} + K\tilde{\bf w} + P{\bf \tilde{d}}
\end{equation}
where ${v} \in \mathbb{R}^{mN}$, $K\in\mathbb{R}^{mN\times h}$, and $P\in\mathbb{R}^{mN\times pN}$ are decision variables to be optimised. Note that, $P$ is an $N\times N$ block matrix with the block size as $m\times p$. Also, the structure of $P$ should ensure the control law is nonanticipative w.r.t. $\tilde{\bf d}$, which is restricted by the fact that only the past prediction error $\tilde{d}_k$ for $k<t$ before time instant $t$ is available for designing $u_t$. As a result, the matrix $P$ should be strictly lower block triangular, which is denoted as $P\in\mathcal{SL}$.

The design objective of flexibility assessment with the control policy for the energy consumption of local RES in \eqref{eq:LDR} can be formulated as following generic form:
\begin{subequations}\label{eq:compact_obj}
\begin{align}
\min\ & J(\gamma_1,\gamma_2) \\
\text{s.t. } & \mathbf{u} \in \mathcal{F}(x_0, \mathbf{w}, \mathbf{d})\\
& \mathbf{u} = v + K\tilde{\mathbf{w}} + P\tilde{\mathbf{d}}
\end{align}
\end{subequations}
where $J(\gamma_{1},\gamma_{2})$ defines the objective for maximizing the amount of flexibility. A simple choice for designing $J(\gamma_1,\gamma_2)$ can be $-\gamma_1-\alpha\gamma_2$, where $\alpha \geq 0$ is a weighting parameter.

Without loss of generality, we can define $\mathcal{X}$ and $\mathcal{U}$ as the following linear constraints:
\begin{equation}\label{eq:state_input_cons}
\mathcal{X} := \left\{\mathbf{x}|G_x{\bf x} \leq g_x\right\},\ 
\mathcal{U} := 
\left\{\mathbf{u}
\left|
\begin{aligned}
& G_u{\bf u} \leq g_u \\
& L_u\mathbf{u} + L_w\mathbf{w} \leq g_{uw}
\end{aligned}
\right.
\right\}
\end{equation}
where $G_x\in\mathbb{R}^{l_x\times nN}$, $G_u\in\mathbb{R}^{l_u\times mN}$, $g_x\in\mathbb{R}^{l_x}$, $g_u\in\mathbb{R}^{l_u}$, $L_{u}\in\mathbb{R}^{l_{uw}\times mN}$, $L_w\in\mathbb{R}^{l_{uw}\times N}$ and $g_{uw}\in\mathbb{R}^{l_{uw}}$ are parameters defined according to the explicit constraints for indoor temperature $\bf x$ and the power from local RES $\bf u$. For the power input $\bf u$, it is constrained from two aspects: one is from the availability of local RES, and another is
the mixed constraints of the power from the grid and the power from local RES, which is induced by the physical limitation of HVAC devices. 

By substituting the control policy \eqref{eq:LDR} into \eqref{eq:lifted} and considering \eqref{eq:grid_power}, the indoor thermal dynamics can be written as
\begin{align}\label{eq:temp}
    {\bf x} =\; & F_x x_0 + F_u v + F_d\hat{\bf d} + F_w{\bf \bar{w}}+ (F_uP + F_d)\tilde{\bf d} \notag\\
    & + (F_u K+F_wM)\tilde{\bf w}.
\end{align}
Then, combining \eqref{eq:flexi_set}-\eqref{eq:temp}, the flexibility assessment can be formulated as the following optimization problem
\begin{subequations}\label{eq:flexi_assessment}
\begin{align}
    \min_{\substack{v,K,P\\\gamma_{1},\gamma_{2}}}\ &J(\gamma_{1},\gamma_{2}) \\
    \text{s.t.}\ &G_x\big(F_x x_0 + F_u v + F_d\hat{\bf d} + F_w{\bf \bar{w}}+ (F_uP + F_d)\tilde{\bf d} \notag\\
    & + (F_u K+F_wM)\tilde{\bf w}\big) - g_x \leq 0,\\
    &G_u\big( v + K\tilde{\bf w} + P\tilde{\bf d} \big) - g_u \leq 0,\\
    &L_u(v + K\tilde{\bf w} + P{\bf \tilde{d}}) + L_w{{\bf \bar{w}} + L_wM\tilde{\bf w}} \notag\\
    & \quad - g_{uw} \leq 0,\\
    & P\in\mathcal{SL},\\
    &\forall \tilde{\bf w}: H_w \tilde{\bf w} - g_w\leq 0,\forall {\bf \tilde{d}}: H_d{\bf \tilde{d}} - g_d\leq 0. \label{eq:universe}
\end{align}
\end{subequations}

Note that the above optimization problem entails considering infinitely many constraints since there are infinite elements of $\tilde{\bf w}$ and $\bf\tilde{d}$ satisfying \eqref{eq:universe}. Consequently, the optimization problem \eqref{eq:flexi_assessment} is semi-infinite, and hence is computationally intractable. In order to make the problem amenable to numerical solutions, we adopt the robust counterpart developed in \cite{bertsimas2022} to reformulate \eqref{eq:flexi_assessment} as a finite-dimensional optimization problem, which results in the following theorem.
}

{\revision\textbf{Theorem 1}: Considering the building thermal dynamics in \eqref{eq:dynamics}, the flexibility description in \eqref{eq:flexi_set}, the prediction uncertainties in \eqref{eq:predict_uncertain}, the control policy in \eqref{eq:LDR}, and the constraints for indoor comfort and control input in \eqref{eq:state_input_cons}, the flexibility assessment problem in \eqref{eq:flexi_assessment} can be reformulated as
\begin{subequations}\label{eq:flexi_final}
\begin{align}
    \min\ & J(\gamma_{1},\gamma_{2})\\
    \text{s.t.}\ &\big[G_x\big(F_xx_0+F_uv+F_d\hat{\bf d}+F_w\bar{\bf w}\big) + g_x\big]_i \notag\\
    & + g_w^{\mathrm{T}}y_1^i + g_d^{\mathrm{T}}y_2^i \leq 0,\\
    & \left[G_x\left( F_uK+F_wM\right)\right]_i = H_w^{\mathrm{T}}y_1^i, \\
    & \left[G_x(F_uP + F_d)\right]_i = H_d^{T}y_2^i,\\
    &\big[G_uv + g_u\big]_j\! +\! g_w^{\mathrm{T}}\mu_1^j + g_d^{\mathrm{T}}\mu_2^j \leq 0,\\
    & \left[G_uK\right]_j = H_w^{\mathrm{T}}\mu_1^j, \\
    & [G_uP]_j = H_d^{\mathrm{T}}\mu_2^j,\label{eq:24g}\\
    & [L_uv+L_w\bar{\bf w}-g_{uw}]_k+g_w^{\mathrm{T}}\eta_1^k + g_d^{\mathrm{T}}\eta_2^k \leq 0,\\
    & [L_uK+L_wM]_k = H_w^{\mathrm{T}}\eta_1^k, \\
    & [L_uP]_k = H_d^{\mathrm{T}}\eta_2^k,\label{eq:24j}\\
    & P\in\mathcal{SL}, \\
    & y_1^i\geq 0, y_2^i\geq 0,\mu_1^j \geq 0, \mu_2^j \geq 0, \eta_1^k\geq 0, \eta_2^k\geq 0
\end{align}
\end{subequations}
where the decision variables are $\gamma_{1}$, $\gamma_{2}$, $g_w$, $v$, $K$, $P$, $y_1^i$, $y_2^i$, $\mu_1^j$, $\mu_2^j$, $\eta_1^k$ and $\eta_2^k$; $i$, $j$ and $k$ are the row indexes of $G_x$, $G_u$ and $L_u$, respectively.}

\textit{Proof}: 
The universal quantifiers in \eqref{eq:universe} can be replaced by considering the following equivalent worst case scenario:\begin{subequations}\label{eq:state_constraints}
\begin{align}
   & G_x\big( F_x x_0\! +\! F_u v\! +\! F_d\hat{\bf d}\! +\! F_w{\bf \bar{w}}\big)
    + \max_{\tilde{\bf w}\in\mathcal{W}}G_x(F_uK \!+\! F_wM)\tilde{\bf w} \notag\\ & +\max_{\tilde{\bf d}\in\mathcal{\tilde{D}}}G_x(F_uP+F_d)\tilde{\bf d} - g_x \leq 0,\label{eq:17b}\\
   & G_u v + \max_{\tilde{\bf w}\in\mathcal{W}}{G_uK\tilde{\bf w}} + \max_{{\bf \tilde{d}}\in\mathcal{\tilde D}} G_uP\tilde{\bf d} - g_u \leq 0, \label{eq:input_cons2}\\
   & L_u v + L_w{\bf \bar{w}} + \max_{\tilde{\bf w}\in\mathcal{W}} \left(L_uK+L_wM\right)\tilde{\bf w}  + \max_{\tilde{\bf d}\in \mathcal{D}}L_uP\tilde{\bf d} \notag\\
   & - g_{uw} \leq 0 \label{eq:17d}
\end{align}
\end{subequations}
where the $\max$ operator denotes row-wise maximization.

Note that all maximization problems for $\tilde{\bf w}$ and $\tilde{\bf d}$ in \eqref{eq:state_constraints} are linear programming for which strong duality holds. So we can apply the duality of LP to obtain their dual problems, which give the same objective value of the primal problems.
For the first maximization term in \eqref{eq:17b}
\begin{subequations}
\begin{align}
\max_{\tilde{\bf w}}\ & \left[G_x\left(F_uK + F_wM\right)\right]_i^{\mathrm{T}}\tilde{\bf w} \\
\text{s.t.}\ &H_w\tilde{\bf w} - g_w\leq 0
\end{align}
\end{subequations}
applying the duality of LP gives the following equivalent dual optimization problem
\begin{subequations}
\begin{align}
    \min_{y_1^i} &\;g_w^{\mathrm{T}}y_1^i\\
    \text{s.t.} &\; \left[G_x\left( F_uK+F_wM\right)\right]_i = H_w^{\mathrm{T}}y_1^i,\quad  y_1^i\geq 0
\end{align}
\end{subequations}
where $y_1^i$ is the vector of Lagrangian multipliers with appropriate dimension. Similarly, for the optimization problem 
\begin{subequations}
\begin{align}
    \max_{\bf {\tilde d}}\ & \left[G_x(F_uP+F_d)\right]_i^{\mathrm{T}}\tilde{\bf d}\\
    \text{s.t.}\ & H_d\tilde{\bf d} -g_d \leq 0
\end{align}
\end{subequations}
its dual problem is 
\begin{subequations}\label{eq:d_x}
\begin{align}
    \min_{y_2^i}\ & g_d^{\mathrm{T}}y_2^i \\
    \text{s.t.}\ & \left[G_x(F_uP + F_d)\right]_i = H_d^{T}y_2^i,\quad y_2^i \geq 0
\end{align}
\end{subequations}
where $y_2^i$ is the vector of Lagrangian multipliers with appropriate dimension.

Following the same arguments, the dual problems for $\max_{{\bf \tilde{w} \in\mathcal{W}}}\left[G_uK\right]_j^{\mathrm{T}}\tilde{\bf w}$, $\max_{\tilde{\bf d}\in\mathcal{\tilde{D}}}[G_uP]_j^{\mathrm{T}}\tilde{\bf d}$ in \eqref{eq:input_cons2} are
\begin{subequations}
\begin{align}
    \min_{\mu_1^j}\ & g_w^{\mathrm{T}}\mu_1^j\\
    \text{s.t.}\ & \left[G_uK\right]_j = H_w^{\mathrm{T}}\mu_1^j,\quad
    \mu_1^j \geq 0
\end{align}
\end{subequations}
and
\begin{subequations}\label{eq:d_u}
\begin{align}
    \min_{\mu_2^j}\ & g_d^{\mathrm{T}}\mu_2^j \\
    \text{s.t.}\ & [G_uP]_j = H_d^{\mathrm{T}}\mu_2^j,\quad
    \mu_2^j \geq 0
\end{align}
\end{subequations}
where $\mu_1^j$ and $\mu_2^j$ are the vectors of Lagrangian multipliers.
The dual optimization problems for 
$\max_{\tilde{\bf w}\in\mathcal{W}}[L_uK+L_wM]_k^{\mathrm{T}}\tilde{\bf w}$ and $\max_{\tilde{\bf d}\in \tilde{D}}[L_uP]_k^{\mathrm{T}}\tilde{\bf d}$ in \eqref{eq:17d} are
\begin{subequations}
\begin{align}
    \min_{\eta_1^k}\ &g_w^{\mathrm{T}}\eta_1^k \\
    \text{s.t.}\ & [L_uK+L_wM]_k = H_w^{T}\eta_1^k,\quad \eta_1^k \geq 0
\end{align}
\end{subequations}
and 
\begin{subequations}\label{eq:mix_dual2}
\begin{align}
\min_{\eta_2^k}\ & g_d^{\mathrm{T}}\eta_2^k \\
\text{s.t.}\ & [L_uP]_k = H_d^{\mathrm{T}}\eta_2^k,\quad \eta_2^k \geq 0
\end{align}
\end{subequations}
respectively, where $\eta_1^k$ and $\eta_2^k$ are multiplier vectors.

The universal quantifiers in \eqref{eq:universe} can be replaced by the optimization problems in \eqref{eq:state_constraints}--\eqref{eq:mix_dual2}. Hence, the semi-infinite optimization problem for flexibility assessment in \eqref{eq:flexi_assessment} can be reformulated as in \eqref{eq:flexi_final}. This completes the proof. \hfill$\square$

After solving the optimization problem \eqref{eq:flexi_final}, the optimal value $\gamma_1^*$ and $\gamma_2^*$ together with the matrix $M$ quantitatively describe the energy flexibility potential of buildings. Then, with this information, the grid operator can generate a feasible DR request $\{\mathcal{R},\mathcal{I}_{se}\}$, which specifies a profile of energy reduction $\mathcal{R}:=[r_0,\cdots,r_h]$ satisfying \eqref{eq:flexi_set} during the time period $\mathcal{I}_{se}:=[t_s,t_e]$ belonging to the prediction horizon, to activate the demand-side flexibility. 

\textit{Remark 2: }Unlike the original optimization problem for flexibility assessment \eqref{eq:flexi_assessment}, which is semi-infinite and computationally intractable, the optimization problem \eqref{eq:flexi_final} has a finite number of decision variables and constraints, which is amenable to numerical solvers. Compared with the original optimization problem, the number of decision variables is increased by $(4h-2)\cdot(l_x+l_u+l_{uw}) + l_d\cdot(l_x+l_u+l_{uw})$, which depends quadratically on the length of the prediction horizon of problem \eqref{eq:flexi_final} and linearly on the length of the flexibility duration. The resulting optimization problem \eqref{eq:flexi_final} is nonconvex since bilinear terms $(g_w^\mathrm{T}y_1^i,g_w^{\mathrm{T}}\mu_1^j,g_w^{\mathrm{T}}\eta_1^k)$ are included in constraints. This type of optimization problem can be dealt with via off-the-shelf solvers, such as Gurobi, SCIP and Ipopt. Warm-starting with the solution of a convex approximation of \eqref{eq:flexi_final} can reduce computation time. In addition, in the case that the ramping rate constraints for energy flexibility are omitted, regularized uncertainty signals of $\tilde{\bf w}$ can be used for control policy design as proposed in \cite{zhang2017robust} to remove the bilinear terms so that the resulting optimization problem only contains convex constraints, which can be solved efficiently.

\textit{Remark 3: }The proposed approach possesses the following properties, which makes it more practical and applicable than existing approaches for unlocking the energy flexibility of buildings:
\begin{itemize}
    \item Compared with the price-based/incentive-based approaches for unlocking buildings' energy flexibility, the proposed scheme can quantitatively describe the maximal admissible set of energy flexibility, even in the face of uncertainties, which is helpful to fully exploit the energy flexibility potential.
    \item Any possible DR requests generated by the grid operator can be achieved robustly without violating the indoor temperature constraints. This property is beneficial for the grid operator to manage its power generation and/or transmission, and possibly different flexibility providers more efficiently.
    \item For dealing with the prediction error of exogenous inputs, unlike scenario-based approaches, the proposed robust optimization-based scheme does not rely on the probability distribution of stochastic prediction errors and only requires its bounds, which is easier to obtain from historical data.
    \item {\revise While the proposed approach entails bidirectional communication between power grid and BMS, the required communication is actually scarce both in frequency and quantity. On the one hand, the communication only happens when the DR event is triggered. On the other hand, only the flexibility capacity and the DR request need to be transmitted. }
\end{itemize}

{\revision
\textit{Remark 4}: Robustly dealing with the estimation error of external conditions via \eqref{eq:d_x} and \eqref{eq:d_u} at each time-step online introduces extra decision variables $(P,y_2^i,\mu_2^i)$ and the corresponding constraints induced.
It should be pointed out that, following a similar idea as Parsi et al \cite{parsi2022computationally}, we can reduce the online computational burden by computing the optimal values of $(P,y_2^i,\mu_2^i)$ offline and keeping them fixed when accessing the energy flexibility online. In this way, the decision variables $(P,y_2^i,\mu_2^i)$, and constraints \eqref{eq:24g} and \eqref{eq:24j} will be removed to improve computational tractability. Since the uncertainty set of external conditions is unchanged, fixing control variables $P$ will not largely degrade control performance.}

\section{Simulation Results}\label{sec:simulation}
\begin{figure}
    \centering
    \includegraphics[width = 0.95\linewidth]{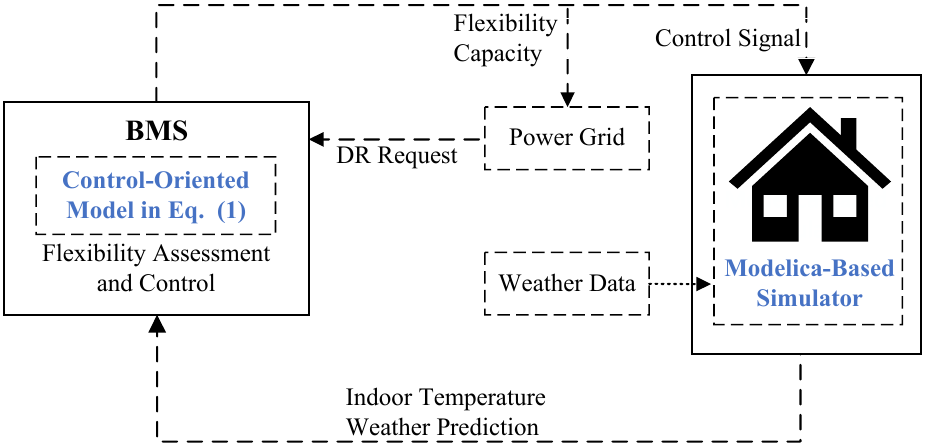}
    \caption{Simulation Diagram.}
    \label{fig:simu_diagram}
\end{figure}

This section explores the effectiveness of the proposed design framework via numerical simulations. In the simulation, all optimization problems are modelled via the {\tt Python} based package {\tt Pyomo} and solved by the {\tt Gurobi} solver. The building model used is {\tt SimpleHouseRad-v0}, which is a {\tt Modelica}-based model, and is adopted from the package {\tt Energym} \cite{scharnhorst2021energym}. For this building model, a heat pump (HP) is used as the heating device. A more detailed model description can be found in \cite{scharnhorst2021energym}. The diagram of our simulation is depicted in Fig. \ref{fig:simu_diagram}.

\begin{algorithm}
\caption{Simulation procedures}
\label{alg:1}
\KwData{weather data, electricity price, training and test data for control-oriented model identification}
\textbf{Select:} prediction model for solar power\\
\While{$t\leq t_{EOS}$}{
Update the nominal power consumption $\bf{\bar{w}}$ every 2 hours (done by BMS)\;
Assess the energy flexibility every 2 hours (done by BMS)\;
Generate a DR request every 2 hours if electricity prices are over 0.15 $\$$/KWh (done by grid operator)\;
Optimize the power drawn from local RES $\bf{u}$ at every 5min sampling instant (done by BMS)\;
$t \leftarrow t+\Delta t$ ($\Delta t$ = 5min)\;
}
\end{algorithm}

\begin{figure}[tbp]
    \centering
    \includegraphics[width = 0.8\linewidth]{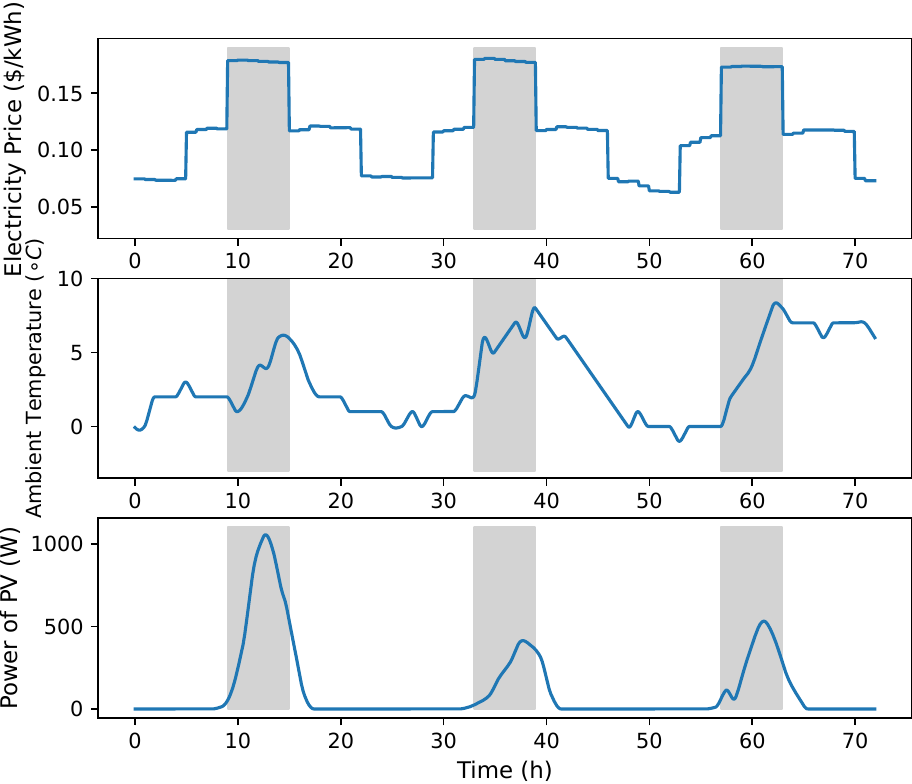}
    \caption{Profiles of electricity prices (top), ambient temperature (middle) and generation by PV panels (bottom).}
    \label{fig:2}
\end{figure}

The detailed procedures of our simulation are summarized in Algorithm \ref{alg:1}, where $t_{EOS}$ represents the end time of simulation and is selected as 72 hours in the simulation. The sampling period is $5$ mins. The weather data is selected as {\tt CH\_VD\_Lausannel}. Solar panels are assumed to provide local RES. The prediction model for available solar power in \cite{mouli2016system} is used, and the size of the solar panel is scaled so that the maximal power output is $1500$W. The profile of electricity price is chosen from PJM (a regional transmission organization that coordinates the movement of wholesale electricity in US). The profiles of electricity price, ambient temperature, and predicted solar power used in the simulation are depicted in Fig.~\ref{fig:2}.

To identify the control oriented model of the building, sensor data, including ambient temperature, indoor temperature, solar radiation, and heating power from HP, over 10 days are used. Then, those data are utilized to determine the RC values via {\tt Scipy}. In our simulation, $\bar{\bf w}$ is computed by minimizing the energy cost while minimizing the deviation of the indoor temperature from its setpoint 21$^\circ$C with the prediction horizon of 12 hours. {\revise Considering the limited flexibility capacity of buildings' thermal inertia, when assessing the energy flexibility, the prediction horizon and activation period of flexibility are all chosen as $2$ hours, which is sufficient to cover the period of short-term DR requests. In addition, for this length of prediction horizon, the flexibility assessment procedure can be finished within a reasonable time.} With the information of the flexibility potential, namely $(\gamma_1^*,\gamma_2^*,M)$, a feasible DR request is sent to the BMS, and the power from RES is updated at each sampling instant by minimizing the difference between the predicted indoor temperature and the indoor temperature setpoint with a prediction horizon of 2 hours.
The scripts for reproducing our simulation results are available in \url{https://github.com/li-yun/Flexibility-Assessment-for-Buildings}.

\begin{table}[tbp]
    \centering
    \caption{Uncertainty sets of exogenous disturbances for different scenarios.}
    \resizebox{\linewidth}{!}{
    \begin{tabular}{cccc}\toprule
         & \multirow{2}{*}{\makecell{uncertainty for ambient\\temperature ($C^{\circ}$)}} & \multirow{2}{*}{\makecell{uncertainty for\\solar radiation ($W/m^2$)}} & \multirow{2}{*}{\makecell{consider the uncertainty\\in flexibility assessment}} \\
         & & & \\\midrule
         scenario 1 & 0 & 0 & Yes\\
         scenario 2 & $[-2, 2]$& $[-50, 50]$& Yes \\
         scenario 3 & $[-5, 5]$& $[-100, 100]$& Yes \\
         scenario 4 & $[-2, 2]$& $[-50, 50]$& No \\
         scenario 5 & $[-5, 5]$& $[-100, 100]$& No \\\bottomrule
    \end{tabular}
    }
    \label{tab:scenarios}
\end{table}
\begin{figure}
    \centering
    \includegraphics[width = 0.9\linewidth]{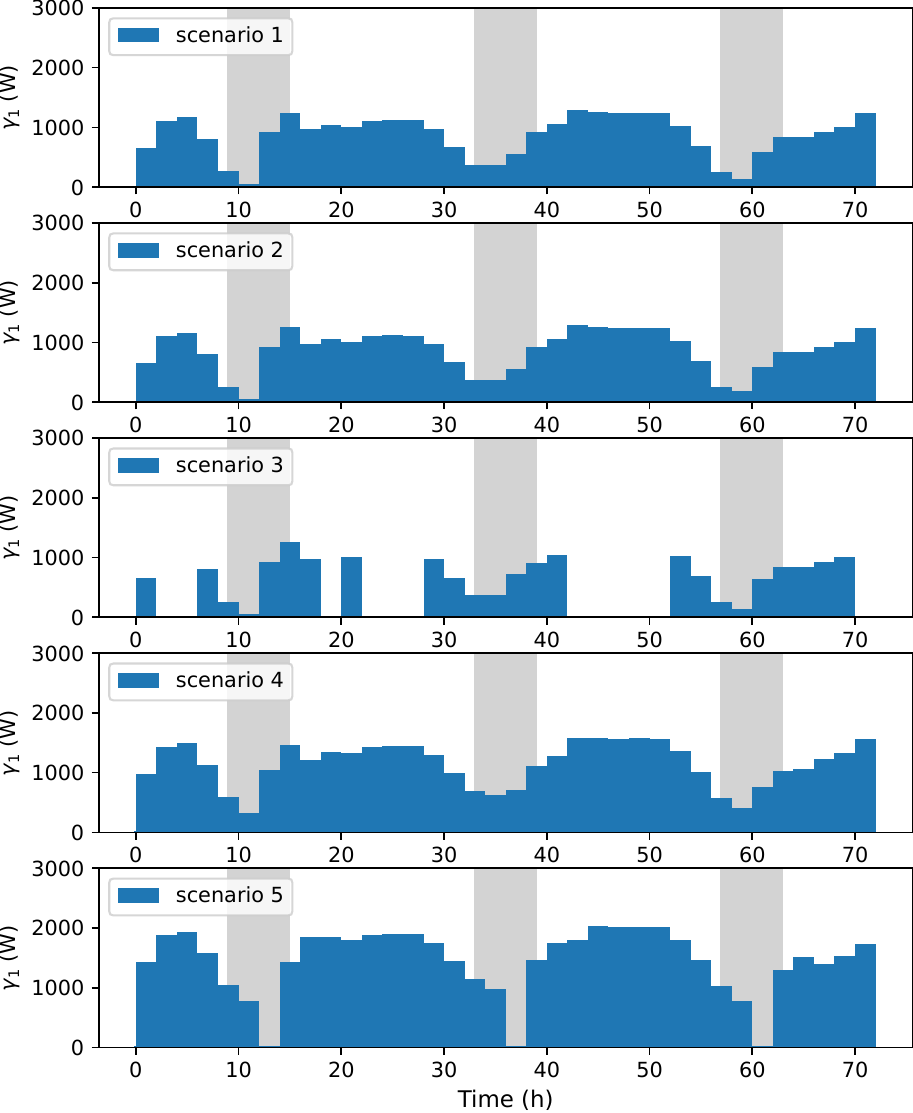}
    \caption{Capacity of flexibility $\gamma_1$ computed via \eqref{eq:flexi_final} for different scenarios.}
    \label{fig:3}
\end{figure}
In order to extensively test the performance of the proposed scheme, we consider five scenarios, whose details are given in Table~\ref{tab:scenarios}. The five scenarios cover three categories: i) no uncertainty in the exogenous input (scenario 1); ii) the exogenous inputs have uncertainty and this uncertainty is also considered in flexibility assessment \eqref{eq:flexi_final} (scenarios 2 and 3); and iii) the exogenous inputs have uncertainty but this uncertainty is not considered in flexibility assessment \eqref{eq:flexi_final} (scenarios 4 and 5). In the simulation, we consider the most challenging uncertainty in our simulation setting: the exogenous inputs are always overestimated with the largest allowable uncertainty. The DR request is generated to request the largest possible reduction of energy consumption when electricity price is over 0.15$\$$/kWh. Under this situation, the BMS tends to overestimate its flexibility potential, and hence risks violating the lower bound of indoor temperature. The comfort range of indoor temperature is set as $[19^\circ \text{C}, 24^\circ \text{C}]$.

\begin{figure}
    \centering
    \includegraphics[width = 0.9\linewidth]{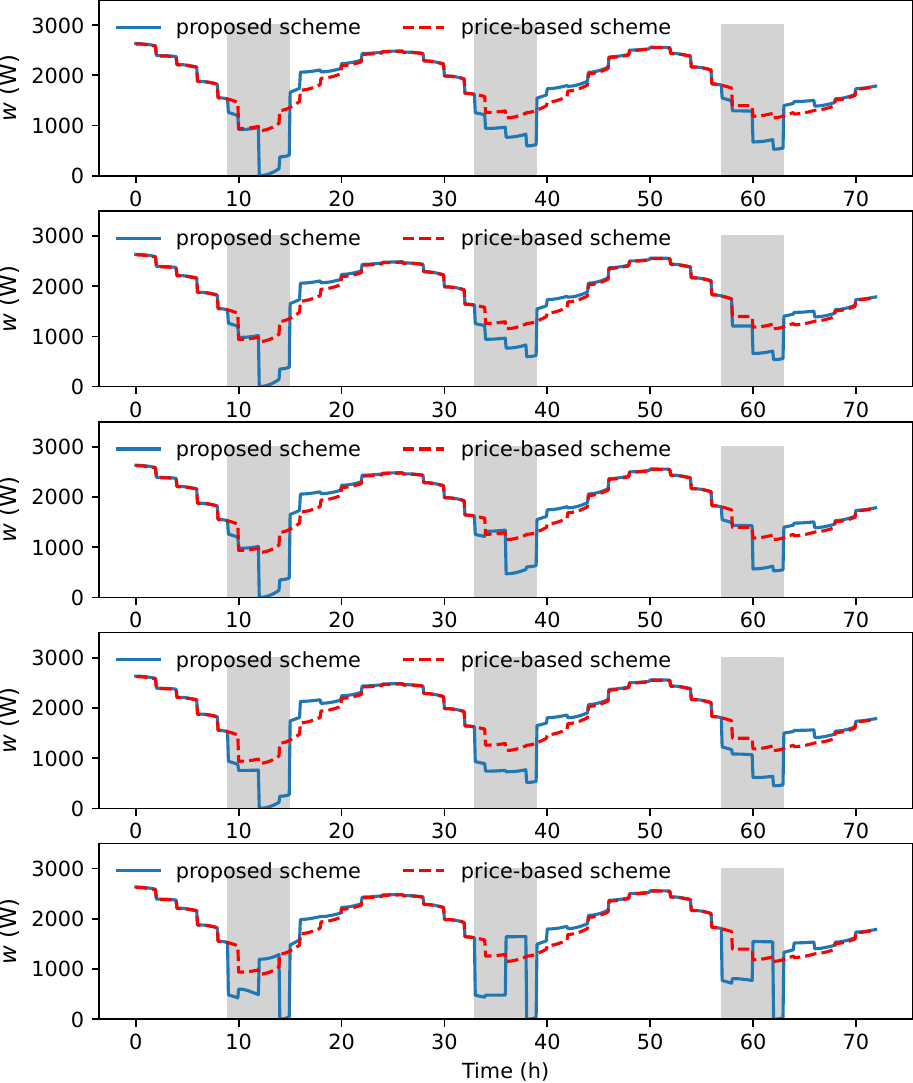}
    \caption{Power consumption from the grid $\bf w$ for different scenarios (Scenarios 1-5 from top to bottom).}
    \label{fig:4}
\end{figure}
\begin{figure}[tbp]
\label{fig:indoor_temp}
    \centering
    \includegraphics[width = 0.9\linewidth]{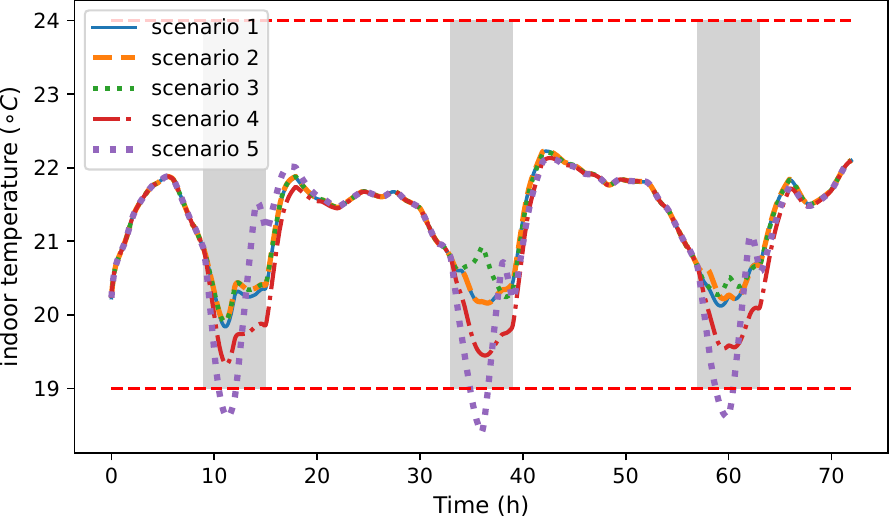}
    \caption{Indoor temperature of different scenarios.}
    \label{fig:5}
\end{figure}

 
Simulation results are depicted in Figs.~\ref{fig:3}--\ref{fig:5}, where the shaded regions with grey colour indicate the period during which DR requests are activated. In Fig.~\ref{fig:3}, the bars filled with blue colour represent the amplitude of flexibility by solving \eqref{eq:flexi_final}. 

It can be observed that for all scenarios the BMS is capable of providing some amount of energy flexibility. For the proposed scheme, with the increase of uncertainties in exogenous conditions, the BMS will become more conservative in assessing its flexibility potential to counteract the uncertainties (Scenarios 1--3). However, without considering the uncertainties in exogenous inputs, the BMS tends to overestimate its flexibility potential (scenarios 4 \& 5).

{\revision Fig. \ref{fig:4} gives the energy consumption profile from the power grid for both our proposed scheme and the conventional price-based scheme, respectively. For the price-based scheme, energy consumption during peak hours is only regulated by increased energy prices. It can be seen that, compared with the price-based approach, our proposed approach achieves more energy reduction during peak hours, which means that more energy flexibility of buildings is unlocked.}

Fig. \ref{fig:5} shows the profile of indoor temperature for different scenarios (red dashed lines indicate the indoor temperature constraints). It is clear that, even in the presence of uncertain exogenous inputs, the proposed design scheme can still achieve the promised energy reduction without violating indoor temperature constraints (Scenarios 1--3). Conversely, without considering the uncertainty of exogenous inputs in the flexibility assessment process, the amount of energy flexibility is overestimated, and hence the indoor temperature constraint can be violated for achieving the promised DR (scenario 5). {\revise In addition, from simulation results we can conclude that the proposed uncertainty description and MPC design is also robust w.r.t. model inaccuracy, which is caused by the discrepancy between the high-fidelity building simulator and the simple linear control-oriented model. }
\section{Conclusion}\label{sec:conclusion}
This work investigates a new design framework for unlocking the energy flexibility associated with buildings' thermal inertia. Unlike the existing price-based/incentive-based approaches for exploiting the energy flexibility of buildings, our proposed approach comprises two steps to provide a unified framework for assessing and exploiting the demand side flexibility of buildings. With the proposed scheme, the energy flexibility of buildings can be quantitatively assessed while considering the uncertainties of external conditions, and the DR requested by the grid operator can also be achieved without violating indoor comfort constraints. The performance of the proposed scheme is verified on a high-fidelity {\tt Modelica} building simulator considering different levels of prediction error for exogenous conditions.

Future extensions may include designing schemes for assessing the flexibility with on-off type heating/cooling devices, and testing the proposed scheme with more complex building models.

\bibliographystyle{IEEEtran.bst}

\bibliography{ref}  
\end{document}